\documentclass[letterpaper,twocolumn,10pt]{article}
\usepackage{usenix-2020-09}
\usepackage{tikz}
\usepackage{amsmath}

\usepackage{filecontents}

\usepackage{kotex}
\usepackage{enumitem}
\usepackage[normalem]{ulem}

\usepackage{booktabs}
\usepackage{multirow}
\usepackage{ulem} 

\usepackage{amssymb}
\usepackage{comment}

\newcommand{\system}{{TempoKV}} 
\newcommand{\devicetitle}{Memory-Semantic Flash} 
\newcommand{\device}{memory-semantic flash}

\usepackage{etoolbox}

\newlength{\TitleBlockLift}
\makeatletter

\patchcmd{\@maketitle}
  {\vbox to 2.5in}
  {\vbox to 2.3in}
  {}
  {\PackageError{title-spacing}
    {Title height patch failed}
    {Check the definition of \string\@maketitle.}}

\patchcmd{\@maketitle}
  {\vskip 2em}
  {\kern-\TitleBlockLift\vskip 2em}
  {}
  {\PackageError{title-spacing}
    {Title lift patch failed}
    {Check the definition of \string\@maketitle.}}

\patchcmd{\@maketitle}
  {\end{center}}
  {\end{center}\kern\TitleBlockLift}
  {}
  {\PackageError{title-spacing}
    {Title compensation patch failed}
    {Check the definition of \string\@maketitle.}}

\makeatother

\begin{document}

\date{}

\title{\Large \bf {\system}: Timely Staging of LLM KV Caches for \devicetitle }

\author{
{\rm Jay H. Park\textsuperscript{†}, Hyungjun Kim, and Dong Kim}\\
\rm {} \\
\textrm{Samsung Semiconductor}\\
} 

\maketitle

\begingroup
\renewcommand\thefootnote{}\footnotetext{%
\textsuperscript{†} Corresponding author: tino.park@samsung.com}
\endgroup

\begin{abstract}
Reusable prefix key--value (KV) caches can outgrow GPU memory in large language model (LLM) serving.
A \device{} hierarchy offers SSD-backed capacity with a limited fast tier, but a logical KV hit is not necessarily ready for GPU retrieval.
Demand staging exposes SSD latency, whereas immediate staging can reserve fast-tier capacity long before retrieval begins.
We present {\it \system{}}, a timing-aware resource-commitment layer that separates early knowledge of reuse from the acquisition of staging resources.
It records reusable-KV hits as metadata-only claims and requests commitment when the runtime-estimated time until retrieval falls to the storage-estimated time needed to make KV resident and protected against eviction.
These estimates adapt to runtime progress and staging state, while commitment remains subject to available protected capacity.
We implement \system{} in vLLM and LMCache on an SSD-backed CXL memory device without changing request scheduling.
Across two models and three prefix cache ratios, \system{} reduces protected fast-tier byte-time per request by 63--91\% versus immediate staging while retaining much of the serving benefit of advance staging.
In a fast-tier capacity sweep, output throughput and p95 time to first token (TTFT) remain nearly unchanged as capacity decreases from 100 to 25~GiB.
Compared with unmodified LMCache's Device-DAX L1 configuration, \system{} reduces p95 TTFT by up to 48.0\% and increases output throughput by up to 27.8\%.

\end{abstract}

\section{Introduction}
\vspace{-0.3cm}
Reusing prefix key--value (KV) state reduces repeated prefill computation in large language model (LLM) serving~\cite{zheng2024sglang,gao2024cost,yu2025pensieve,wang2025kvcache}.
As reusable KV outgrows GPU high-bandwidth memory (HBM), systems retain it in host memory and SSD-backed storage~\cite{yu2025pensieve,qin2025mooncake,chen2025impress,hu2026bidaw,shin2026matkv}, but retrieval can incur substantial latency~\cite{liu2024cachegen,chen2025impress,gao2025hcache}.
We focus on prefix KV retained between requests, rather than live KV accessed during generation.
We study this reuse in \emph{\device{}}, a storage architecture combining SSD-backed capacity with a smaller fast-memory tier behind a memory-oriented abstraction.
SSD-backed CXL memory realizes this architecture using device DRAM as its fast tier~\cite{zhang2025skybyte,jang2026itme,jang2026hymcache}.
Yet a logical KV hit may still require SSD-to-fast-tier staging before the KV can be retrieved into GPU memory.

The challenge is when to commit staging resources.
Commitment authorizes staging and reserves sufficient fast-tier capacity to protect matched KV from eviction until its GPU transfer completes.
Waiting until retrieval is requested exposes SSD staging latency; committing at hit discovery can tie up capacity long before retrieval, limiting ordinary caching and other staging work.
Existing systems use scheduler look-ahead, queued requests, or application workflows to prepare reusable KV in advance~\cite{gao2024cost,xie2026strata,agarwal2026symphony,pan2026kvflow}.
Queue rank, for example, indicates order rather than time until retrieval.
The available lead depends on runtime progress, whereas the required staging time depends on KV residency, footprint, outstanding staging work, and the effective staging rate.
Advance notice of reuse creates an opportunity to hide SSD latency, not a requirement to reserve fast-tier capacity immediately.
The goal is to exploit that lead time without holding protected capacity longer than necessary.

We present {\it \system{}}, a timing-aware re\-source-com\-mit\-ment layer that separates early knowledge of reuse from staging commitment.
It records identified reusable-KV hits for queued requests as metadata-only claims without initiating I/O or reserving capacity.
The runtime estimates \emph{time-to-use} (TTU), the time until fast-tier-to-GPU retrieval begins, while a storage-side staging provider estimates \emph{time-to-ready} (TTR), the time needed to make matched KV resident and protected if committed now.
The controller requests commitment when TTU is at or below TTR and reevaluates timing as runtime and staging state change; commitment remains subject to the provider's validity and capacity checks.

We implement \system{} by extending vLLM~\cite{kwon2023vllm} and LMCache~\cite{liu2025lmcache}, with the provider for an SSD-backed CXL Type-3 memory device.
The integration preserves request scheduling and reuses existing prefix-lookup and GPU-transfer paths without changes to device firmware or the CXL data path.
Across two models and three prefix cache ratios, \system{} reduces protected fast-tier byte-time per request by 63--91\% compared with immediate staging while retaining much of the serving benefit of advance staging.

\vspace{-0.2cm}
\section{Background and Motivation}
\label{sec:background}
\vspace{-0.2cm}

\subsection{Reusable KV on \devicetitle{}} 
\label{sec:bg-reusable-kv} 
\vspace{-0.2cm}

Unlike live KV accessed throughout generation, retained prefix KV remains quiescent until a later request matches it, making it suitable for a high-capacity backing tier below HBM.
A \device{} hierarchy combines SSD-backed capacity with a smaller DRAM or other fast-memory tier behind a memory-oriented abstraction.
SSD-backed CXL memory is one realization, but the staging-timing problem does not depend on CXL-specific transport semantics.
A KV object is an independently identified unit of reusable prefix KV.
Reusable KV is read-only during reuse, and once a prefix match is known, the required KV objects and their access order can be identified before retrieval begins.
Beluga and TraCT use CXL memory pools as shared KV-cache substrates~\cite{yang2026beluga,yoon2025tract}.
More directly, ITME exposes software-directed prefetching into the internal DRAM cache of SSD-backed CXL-hybrid memory~\cite{jang2026itme}, while HyMCache streams reusable prefix KV through a bounded internal-DRAM window~\cite{jang2026hymcache}.
Our focus is when to commit staging resources for a waiting request, using runtime lead time and storage-side preparation time.

\begin{figure}[t]
    \centering
    \includegraphics[scale=0.45]{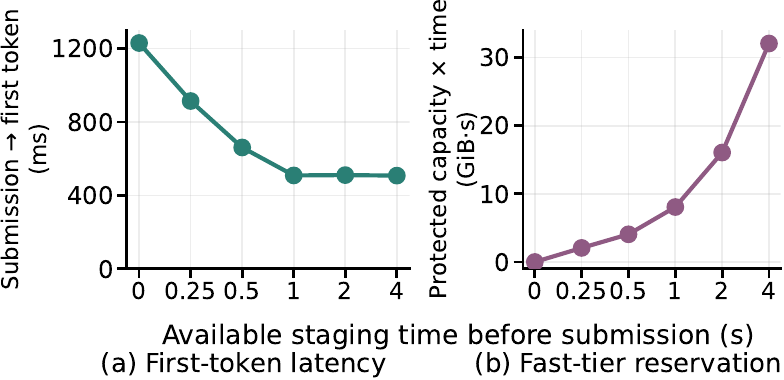} 
    \vspace{-0.3cm}
    \caption{Earlier staging hides SSD staging latency but increases protected fast-tier byte-time.}
    \vspace{-0.2cm}
    \label{fig:staging-timing}
\end{figure}

\subsection{Memory Exposure Is Not Readiness}
\label{sec:bg-readiness} 
\vspace{-0.2cm}

A logical KV hit does not establish fast-tier residency: required objects may still need SSD-to-fast-tier staging.
Time to readiness depends on queued and in-progress staging work, remaining bytes, and the effective staging rate.
We distinguish SSD-to-fast-tier staging from fast-tier-to-GPU retrieval.
Storage-side logic controls staging and protection, whereas the runtime determines when retrieval begins.

We designate the matched KV as \emph{stage-ready} once every required KV object is present in the fast tier and protected against eviction.
\system{} uses this boundary so that the existing fast-tier-to-GPU retrieval path can proceed without further SSD staging, at the cost of protecting the full matched KV footprint.
We use \emph{staging commitment} for the point at which staging is authorized and sufficient fast-tier capacity is reserved to keep the matched KV protected through retrieval.

\subsection{The Staging-Timing Problem}
\label{sec:bg-timing}
Prior systems prefetch reusable KV for queued requests~\cite{gao2024cost,wang2026pcr,xie2026strata}, while others derive advance signals from user interaction or application workflows~\cite{agarwal2026symphony,pan2026kvflow}.
These mechanisms provide advance notice of future reuse, but such notice alone does not establish a reliable wall-clock trigger for staging.
Unlike Bidaw's storage-aware request scheduling and KV-read ordering~\cite{hu2026bidaw}, we time staging commitment without changing runtime request scheduling.

\textbf{Early staging trades latency for capacity.}
Figure~\ref{fig:staging-timing} varies the staging lead---the time from staging initiation to request submission---for an SSD-resident, 64k-token Llama-3.1-8B-Instruct~\cite{grattafiori2024llama3} prefix with an 8-GiB KV footprint.
Before each condition, we retain the KV on SSD and clear the fast tier; zero lead starts staging at submission.
Figure~\ref{fig:staging-timing}(a) measures submission-to-first-token latency, excluding the lead;
Figure~\ref{fig:staging-timing}(b) measures protected fast-tier byte-time from commitment to submission.
Median latency falls from 1.23 to 0.51~s as lead increases from 0 to 1~s, then plateaus.
Extending the lead to 4~s provides no additional latency benefit but raises protected fast-tier byte-time from 8 to 32~GiB$\cdot$s.
Staging should therefore begin early enough to hide SSD staging latency; beyond that point, additional lead increases protected fast-tier byte-time without further reducing latency.

\begin{figure}[t]
    \centering
    \includegraphics[scale=0.33]{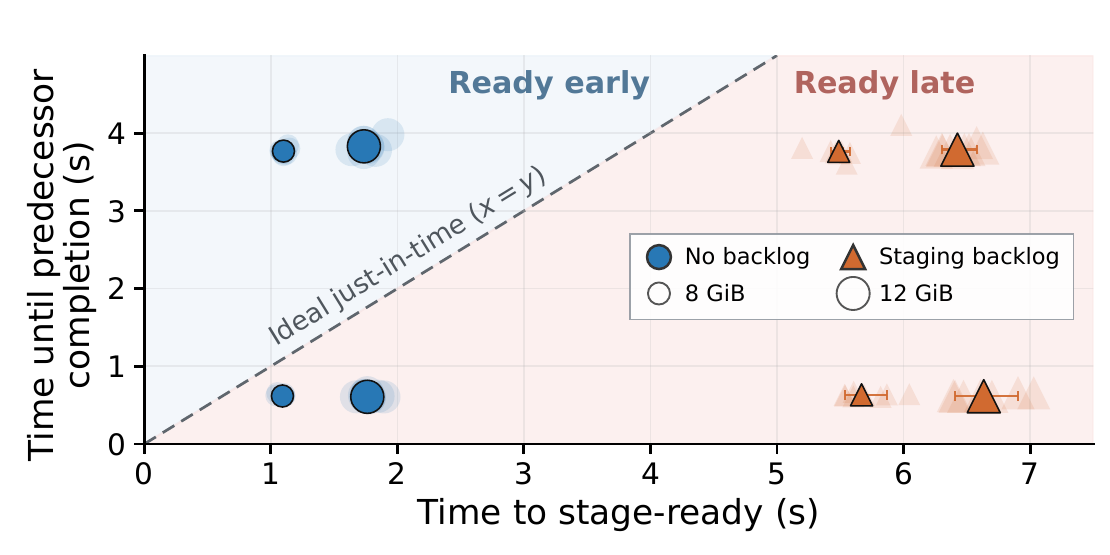} 
    \vspace{-0.5cm}
    \caption{Queue rank is not a staging deadline.}
    \vspace{-0.3cm}
    \label{fig:rank-timing}
\end{figure}

\textbf{Queue rank is not a staging deadline.}
Figure~\ref{fig:rank-timing} tests a rank-one trigger under FCFS with one active sequence: each target is the sole waiter behind a running predecessor.
We vary the predecessor's nominal remaining service time (0.5 or 4~s), the target KV footprint (8 or 12~GiB), and staging backlog (none or four committed 8-GiB operations ahead of the target).
Protected capacity is sufficient, so backlog does not introduce capacity waits.
The dashed $x=y$ line marks just-in-time completion; $y-x$ is the timing margin (positive is early; negative is late).
Despite identical rank, margins range from 2.67~s early to 6.02~s late; with a 4-s predecessor, adding backlog makes both target sizes switch from early to late.

Together, Figures~\ref{fig:staging-timing} and \ref{fig:rank-timing} show why staging must be timed and why queue rank alone is not a sufficient timing signal.
Timely commitment therefore requires comparing the runtime-estimated time until retrieval begins with the storage-estimated time to reach stage-ready, subject to available protected capacity.
The goal is to make matched KV stage-ready by the expected start of retrieval without reserving fast-tier capacity unnecessarily early.

\section{\system{} Design}

\system{} is a timing-aware resource-commitment layer for SSD-to-fast-tier staging in a \device{} hierarchy.
Given a waiting request with identified reusable KV objects, it determines when to request staging commitment.
It leaves prefix matching and the runtime's request-scheduling policy unchanged and reuses the fast-tier-to-GPU retrieval path.

\begin{figure}[t]
    \centering
    \includegraphics[scale=0.55]{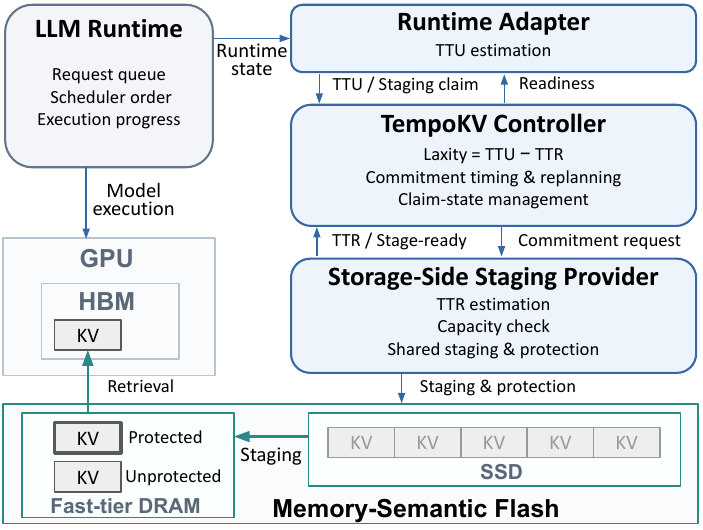}
    \vspace{-0.3cm}
    \caption{\system{} architecture overview. The controller combines runtime-estimated TTU and provider-estimated TTR to time staging commitment requests.}
    \vspace{-0.3cm}
    \label{fig:tempokv-architecture}
\end{figure}

Figure~\ref{fig:tempokv-architecture} shows \system{}'s three logical components:
the runtime adapter, the \system{} controller, and the storage-side staging provider.
The runtime adapter estimates when retrieval is expected to begin;
the provider estimates the time required to reach stage-ready and controls staging resources; 
and the controller combines these estimates to time commitment.
The runtime need not inspect storage internals, and the provider need not reconstruct request execution.
Here, storage-side denotes responsibility for staging state and fast-tier resources, not a requirement that the provider execute inside the device.

The runtime adapter reports each identified reusable-KV hit to the controller as a request-scoped record, called a \emph{staging claim}.
The claim contains an ordered manifest of the required KV objects and carries a revisable estimate of the time remaining until retrieval begins.
Before commitment, reporting or revising a claim neither initiates staging I/O nor reserves protected fast-tier capacity.

\subsection{Two-Sided Timing}
\label{sec:design-architecture}

\textbf{Runtime-estimated time-to-use (TTU).}
The runtime determines the available lead time: TTU estimates how long remains until fast-tier-to-GPU retrieval is expected to begin, not until the model consumes the KV or
produces its first token.
The runtime adapter uses the active batch, per-request execution progress, and the runtime's scheduling order.
The projection assumes the target's matched KV is stage-ready.
Otherwise, the target's unfinished staging would postpone its predicted retrieval start and thereby defer the very commitment needed to prepare it.

The adapter makes a read-only projection of running requests and queued requests ahead of the target, mapping unfinished prefill and expected decode work to time using calibrated execution costs.
It accounts for resources released as requests complete and for preceding requests' known retrieval dependencies.
Concurrent work advances on a shared timeline under the runtime's scheduling order and resource constraints, rather than summing overlapping request latencies.
Output-length estimates come from recent completed requests.
For a running request with $n$ generated tokens, remaining output is the mean of $N-n$ over completed outputs of length $N>n$; the generation limit bounds the predicted total length.
The projected time until the target's retrieval starts gives TTU, which is revised as runtime state changes.
This estimator adapts work-based waiting-time estimation from QLM~\cite{patke2024qlm} and output-length conditioning from Past-Future~\cite{gong2025pastfuture}.

\textbf{Provider-estimated time-to-ready (TTR).}
The provider estimates the preparation time behind a logical KV hit: TTR predicts how long the matched KV would take to become resident and protected if committed now.
The estimate uses known fast-tier residency, queued and in-progress staging, and the effective staging service rate.
It excludes pre-commitment waits and subsequent fast-tier-to-GPU retrieval; capacity availability is checked separately.

The provider evaluates a hypothetical commitment through a metadata-only projection of its staging queue.
It adds SSD reads only for required objects neither known to be resident nor covered by existing staging.
Objects covered by queued or in-progress staging retain those completion dependencies; resident but unprotected objects require protection, not reads.
The projection follows the provider's issue order and concurrency.
Concurrent operations share an effective aggregate service rate calibrated and updated from completed operations over active staging intervals.
TTR is determined by when the last required object becomes resident and protected, rather than by summing overlapping transfer times.
Thus, a claim can add no new SSD reads yet still have a nonzero TTR while waiting for shared staging.
The provider revises TTR as staging progresses and service conditions change, accounting for contention in line with load-aware informed prefetching~\cite{tomkins1997informed}.

The timing interface requires revisable TTU and TTR estimates with the meanings defined above, but does not prescribe a particular estimation algorithm.
Either estimator can be replaced without changing the commitment mechanism.

\vspace{-0.2cm}
\subsection{Timely Commitment and Replanning}
\label{sec:design-commitment}
\vspace{-0.1cm}

\textbf{Timing-based eligibility.}
Using the latest estimates at control time $t$, the controller computes the estimated \emph{laxity} of an uncommitted claim $c$ as $\widehat{L}_c(t) \triangleq \widehat{\mathrm{TTU}}_c(t) - \widehat{\mathrm{TTR}}_c(t)$.
Positive, zero, and negative laxity predict early, just-in-time, and late readiness, respectively, if committed now.
A claim becomes eligible when $\widehat{L}_c(t)\leq 0$ or the runtime requests retrieval.
Eligibility permits a commitment attempt; successful commitment remains subject to the provider's validity and capacity checks.

\textbf{Commitment and replanning.}
Eligibility does not itself reserve fast-tier capacity.
The controller submits an eligible claim to the provider, which rechecks KV-object validity, fast-tier residency, and the additional protected capacity required before committing the claim.
A protected commitment requires the claim's full KV footprint to fit within the configured protected-capacity budget.
If sufficient capacity is unavailable, the claim remains in its initial \textsc{Planned} state, without staging protection or reserved fast-tier capacity.
Timing therefore guides commitment attempts, while the provider determines whether commitment is feasible now.

The controller prioritizes claims with pending runtime retrieval requests and considers other eligible claims in the latest request order reported by the runtime.
An ineligible or capacity-blocked claim does not block commitment attempts for other claims.
Commitment order may therefore differ from runtime request order:
a claim with a longer TTU can become eligible earlier if its TTR is sufficiently larger.
These decisions govern staging commitment without changing the runtime's request-scheduling policy.

Until commitment, the controller reevaluates claims on request-order and timing-estimate updates, staging progress, and capacity release, as well as on periodic control ticks.
It adjusts the latest timestamped TTU for elapsed time and uses the provider's latest TTR estimate, which reflects current staging state rather than elapsed time alone.
Before the runtime requests retrieval, revised estimates may make an uncommitted claim's laxity positive and defer further commitment attempts.
Once retrieval is requested, the claim remains eligible until it is committed or the ordinary retrieval path is selected.
After each successful commitment, subsequent decisions use the provider's updated staging and capacity state.
Once a claim is committed, request-order and timing-estimate updates do not revoke its commitment or preempt its staging work, avoiding wasted I/O and repeated resource acquisition.

\textbf{Protected capacity and on-demand retrieval.}
Protected staging and ordinary caching share the fast tier.
The provider enforces a protected-capacity budget below total capacity, counting both protected objects and reservations for outstanding staging work without statically partitioning memory.

If an uncommitted claim cannot obtain protection when the runtime requests retrieval, the request proceeds through the existing demand retrieval path.
The controller makes this choice mutually exclusive with commitment.
Once the ordinary path is selected, delayed updates cannot commit the same claim.
Ordinary retrieval preserves KV reuse but does not provide the stage-ready guarantee: nonresident data is fetched from SSD on demand.
Ordinary retrieval does not reserve protected capacity, so exhaustion of the protected-capacity budget alone does not block this path.
Claims that have already committed retain their protection and reach stage-ready before retrieval.
Uncommitted requests therefore need not wait for a protected staging commitment, although ordinary retrieval may still incur SSD latency and contention.

\textbf{Shared staging and resource lifecycle.}
Claims are request-specific, but their staging work and protection may be shared.
The provider coalesces staging for the same KV object and reserves additional fast-tier capacity only for bytes not already covered by another committed claim's protection or staging reservation.
For each remaining object, it either protects an existing fast-tier copy or stages the object from SSD and protects it.
Sharing therefore avoids duplicate SSD reads and duplicate capacity reservations, although a newly committed claim may prolong an object's protection.

Commitment moves a claim from \textsc{Planned} to \textsc{Staging}, or to \textsc{StageReady} if every required object is resident and protected at commitment.
Otherwise, the claim becomes \textsc{StageReady} only after the provider confirms every required object is resident and protected.
The provider reports readiness to the controller, which updates the claim's state and returns it in response to the runtime adapter's readiness queries.
The claim enters \textsc{Transferring} only when the runtime begins retrieval, not merely when readiness is reported.

Protection remains in force until the claim's GPU retrieval completes.
At release, the provider ends protection on behalf of the claim.
Capacity becomes reusable only for bytes no longer protected on behalf of another committed claim.
\textsc{Released} ends this claim's protection guarantee, not physical residency: unprotected objects may remain as evictable cache entries.
An uncommitted claim can be cancelled immediately.
A committed claim can be cancelled, but protection on its behalf is retained until outstanding staging I/O and GPU accesses can no longer use the associated objects.

\begin{figure*}[t]
    \centering
    \includegraphics[width=\textwidth]{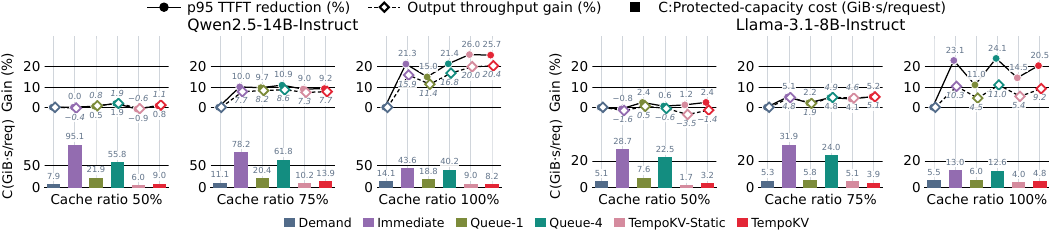}
    \vspace{-0.8cm}
    \caption{Performance and protected-capacity cost at prefix cache ratios of 50\%, 75\%, and 100\%. Upper panels show p95 TTFT reduction and output-throughput gain versus Demand (higher is better); lower panels show $C$ (lower is better).}
    \vspace{-0.3cm}
    \label{fig:eval-overlay}
\end{figure*}

\section{Implementation}
\label{sec:implementation}
\textbf{Runtime and controller integration.}
We implement \system{} by extending vLLM~\cite{kwon2023vllm}
and LMCache~\cite{liu2025lmcache}.
After each scheduling iteration, a runtime adapter computes TTU from a bounded scheduler snapshot and sends only claim metadata to the controller in LMCache's multiprocess service.
The adapter queries the controller for claim readiness.
If an uncommitted claim cannot obtain protection when retrieval is requested, the adapter coordinates with the controller to select ordinary retrieval instead.
The integration preserves vLLM's request-scheduling policy and reuses LMCache's prefix-lookup and GPU-transfer paths.

\textbf{Storage-side staging provider.}
We implement the provider for a memory-semantic flash device using XCENA MX1P~\cite{xcenainfinitememory} and integrate it into LMCache through a DAX-backed L1 cache.
MX1P realizes the \device{} architecture as a CXL Type-3 memory device with an SSD backing tier and device DRAM as its fast tier.
The provider maps KV manifests to device-memory ranges and uses the device's existing range-management APIs for staging and protection.
It estimates TTR from known residency, tracked staging work, and the observed effective staging rate.
It protects required ranges with \texttt{pin} and releases protection with \texttt{unpin} only when no committed claim, outstanding staging I/O, or GPU access requires it.
Staging-completion callbacks update the controller.
The provider enforces the configured protected-capacity budget within the device's pin limit, without statically partitioning the fast tier.
 
\vspace{-0.3cm}
\section{Evaluation}
\label{sec:evaluation}
\vspace{-0.3cm}
We evaluate whether \system{} reduces protected-capacity cost while maintaining or improving serving performance.
We compare commitment policies and a static-TTR ablation, examine sensitivity to fast-tier capacity, and compare against LMCache's Device-DAX L1 configuration.

\textbf{Experimental setup.}
We use a 72-core Intel Xeon Granite Rapids server with one NVIDIA H100 PCIe GPU (80~GB HBM) connected over PCIe Gen5$\times$16.
A \device{} device has 128~GiB of DRAM and is backed by a 15.36~TB Samsung PM1753 NVMe SSD connected over PCIe Gen5$\times$4.
The device connects to the host over CXL$\times$8.
Unless otherwise stated, the policy experiments use a 100~GiB device-DRAM fast tier with a protected-capacity budget of half that capacity.
The device applies LRU eviction to unprotected entries in the shared fast tier.

These experiments use modified vLLM~v0.23.0 and LMCache~v0.5.1, with Qwen2.5-14B-Instruct~\cite{qwen2025qwen} and Llama-3.1-8B-Instruct~\cite{grattafiori2024llama3} in BF16 for both inference and KV caches.
Qwen uses YaRN with a scaling factor of four.
We use vLLM's FCFS with up to four active sequences and GPU prefix caching disabled.
The controller combines event-driven updates with a 50~ms periodic tick.

\textbf{Metrics.}
We report p95 time to first token (TTFT), measured from client submission to receipt of the first output token.
Output throughput (token/s) is the total output tokens in successful responses divided by the time from the first submission to the last response completion.
For the policy experiments, \emph{protected-capacity cost} is $C=(\int_0^T P(t)\,dt)/N$, in GiB$\cdot$s/request, where $P(t)$ is reserved fast-tier capacity in GiB and $N$ is the number of completed requests.
The interval $[0,T]$ extends from the first submission until the run has completed and all staging reservations have been safely released.
$P(t)$ includes capacity reserved for unfinished staging and counts shared KV objects only once;
$C$ measures reservation, not physical DRAM occupancy.

\subsection{Performance and Protected-Capacity Cost}
\label{sec:eval-policy}
Figure~\ref{fig:eval-overlay} compares six configurations across models and three prefix cache ratios under the default fast-tier configuration.

\textbf{Workload.}
We use 12 full NarrativeQA~\cite{kovcisky2018narrativeqa} documents, four from each of three length groups, with one fixed question per document.
Both models receive the same documents, questions, and request order; input lengths span 16,078--67,744 tokens across the two tokenizers.
We vary the nominal prefix cache ratio, defined relative to each document's token count, across 50\%, 75\%, and 100\%.
We retain KV for the corresponding prefix in complete 256-token LMCache chunks while submitting unchanged full prompts.
Even at a 100\% prefix cache ratio, the question and any final incomplete chunk remain uncached.
A higher ratio reduces uncached prefill but increases the matched KV footprint, varying runtime work relevant to TTU and staging demand relevant to TTR.

Before each run, we retain precomputed KV on SSD, reset the fast tier, and initialize predictors from the same model-specific calibration.
Requests arrive at fixed 0.5-s intervals, independently of response completion, without cache resets between requests.
Generation is capped at 128 output tokens with natural end-of-sequence termination.

\textbf{Policies.}
The six configurations share validity and capacity checks and the staging, retrieval, and release paths of the same runtime--controller--provider implementation.
\textbf{Demand} makes a claim eligible when the runtime requests retrieval.
\textbf{Immediate} does so when the adapter reports the reusable-KV hit to the controller.
\textbf{Queue-$K$} does so when the request is among the first $K$ in the runtime's full waiting queue, for $K\in\{1,4\}$.
\textbf{\system{}-Static} and \textbf{\system{}} use the same TTU estimator and the rule $\widehat{L}_c(t)\leq 0$ (Section~\ref{sec:design-commitment}).
Retrieval demand makes any claim eligible; an uncommitted claim unable to obtain protection uses ordinary retrieval.

\system{}-Static estimates each claim's TTR in isolation using the current residency and protection of its required KV and a premeasured, model-specific staging rate held fixed during evaluation (approximately 11~GiB/s).
It excludes staging queueing, contention, and ongoing staging progress, as well as runtime-measured overheads in reaching stage-ready.

\textbf{Performance--capacity tradeoff.}
Figure~\ref{fig:eval-overlay} shows that \system{} retains much of the serving benefit of advance staging while substantially reducing its protected-capacity cost.
Across all six model--ratio combinations, it reduces $C$ by 63--91\% relative to Immediate and 62--86\% to Queue-4.
The cost of early commitment is visible at a 50\% prefix cache ratio, where serving gains are small despite high reservation cost.
At 100\%, \system{} reduces p95 TTFT by up to 25.7\% and increases output throughput by up to 20.4\% over Demand.

A smaller queue-rank threshold does not achieve the same balance: at 100\%, \system{} outperforms Queue-1 on both serving metrics at lower $C$ in both models.
For Qwen at this ratio, it also outperforms Immediate on both serving metrics.
For Llama, Queue-4 is faster, but \system{} trades 4.7\% higher p95 TTFT and slightly lower throughput
for 62\% lower $C$.
These results favor comparing the estimated time until retrieval with the estimated time to stage-ready, rather than using hit discovery or queue rank alone to trigger commitment.

\textbf{Effect of state-aware TTR estimation.}
Timing-based commitment remains effective with a simple TTR estimate, while storage-state awareness can improve serving performance further.
For Qwen at 75\% and 100\% prefix cache ratios, \system{}-Static already captures most of the serving gains.
For Llama at 100\%, state-aware TTR increases the p95 TTFT reduction over Demand from 14.5\% to 20.5\% and improves throughput, while $C$ rises from 4.0 to 4.8~GiB$\cdot$s/request.
Here, accounting for storage state improves serving performance at some additional protection cost, rather than simply minimizing $C$.

\begin{figure}[t]
    \centering
    \includegraphics[scale=0.58]{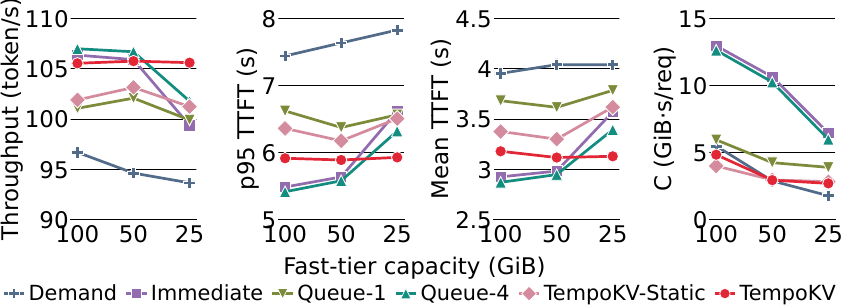}
    \vspace{-0.3cm}
    \caption{Sensitivity to fast-tier capacity for Llama-3.1-8B-Instruct at a 100\% prefix cache ratio.}
    \vspace{-0.3cm}
    \label{fig:eval-capacity}
\end{figure}

\vspace{-0.3cm}
\subsection{Sensitivity to Fast-Tier Capacity}
\label{sec:eval-capacity}
\vspace{-0.2cm}
We reuse the Llama workload from Section~\ref{sec:eval-policy} at a 100\% prefix cache ratio and compare all six configurations.
The fast tier is configured to 100, 50, and 25~GiB, with corresponding protected-capacity budgets of 50, 25, and 12.5~GiB.

Figure~\ref{fig:eval-capacity} shows that \system{} sustains serving performance despite a 75\% reduction in fast-tier capacity: throughput and p95 TTFT remain nearly unchanged at approximately 106~token/s and 5.9~s, respectively.
The latency advantage of Immediate and Queue-4 over \system{} at 100~GiB reverses at 25~GiB.
At this capacity, \system{} achieves the highest throughput and lowest mean and p95 TTFT among all six configurations, with lower $C$ than Immediate and Queue-4.
Controller logs show that insufficient protection budget defers commitment for only one request under \system{}, compared with six under each of Immediate and Queue-4.
This contrast is consistent with early reservations constraining subsequent staging when protected capacity is scarce.

Low reservation cost alone does not yield the same performance: at 25~GiB, \system{}-Static incurs similar $C$ but delivers lower throughput and higher TTFT than \system{}.
Together, these results show that commitment timing matters not only for reservation cost, but also for sustaining serving performance under capacity pressure.
In this workload, \system{} retains nearly the same throughput and p95 TTFT with one-quarter of the default fast-tier capacity and protected-capacity budget.

\begin{figure}[t]
    \centering
    \includegraphics[scale=0.45]{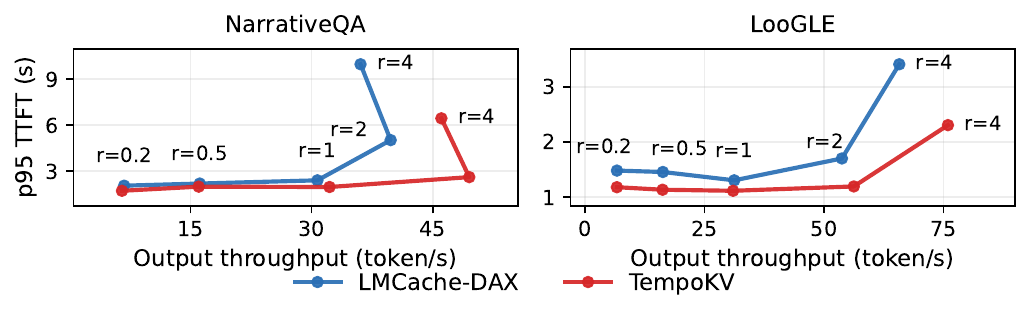} 
    \vspace{-0.4cm}
    \caption{End-to-end serving performance versus LMCache-DAX. Each point corresponds to an offered request rate $r$.}
    \vspace{-0.5cm}
    \label{fig:ttft_vs_throughput}
\end{figure}

\vspace{-0.3cm}
\subsection{Comparison with the DAX-L1 Baseline}
\label{sec:eval-dax-baseline}
\vspace{-0.2cm}
We compare \system{} with \textbf{LMCache-DAX}, which maps the device's exposed memory as L1 using LMCache's Device-DAX L1 configuration~\cite{lmcachempconfig}.
Unlike Demand, this baseline runs unmodified LMCache without our runtime adapter, controller, or staging provider.

Both systems use Llama-3.1-8B-Instruct with a 100\% prefix cache ratio and a 100~GiB fast tier; \system{} enforces a 50~GiB protected-capacity budget.
We randomly sample 16 contexts from each of NarrativeQA and LooGLE~\cite{li2024loogle}, using one associated question per context.
The sampled requests and their order are fixed across both systems and all request rates.
Generation is capped at 32 output tokens.
For both workloads, requests arrive at fixed intervals of $1/r$~s, independently of response completion, with the offered request rate $r$ ranging from 0.2 to 4~requests/s.
Before each run, we clear the fast tier while retaining precomputed KV on SSD.

Figure~\ref{fig:ttft_vs_throughput} shows that \system{} improves the end-to-end latency--throughput tradeoff over LMCache-DAX on both workloads.
At matched offered request rates, \system{} reduces p95 TTFT by up to 48.0\% and increases output throughput by up to 27.8\% on NarrativeQA.
The corresponding improvements on LooGLE reach 32.4\% and 15.5\%, respectively.

\vspace{-0.6cm}
\section{Conclusion}
\vspace{-0.3cm}
\system{} separates early knowledge of KV reuse from staging-resource commitment in \device{} hierarchies.
By comparing runtime-estimated time-to-use with provider-estimated time-to-ready, it times commitment for SSD-to-fast-tier staging while preserving the runtime's request-scheduling policy.
Our evaluation on SSD-backed CXL memory shows that \system{} retains much of the serving benefit of advance staging at substantially lower protected-capacity cost than immediate staging.

\bibliographystyle{plain}
\bibliography{references}

\end{document}